**Boron Nitride Nanosheets for Metal Protection**

*Lu Hua Li,[†]\* Tan Xing,[†] Ying Chen[†]\* and Rob Jones[‡]*

Dr. L. H. Li, X. Tan, Prof. Y. Chen
Institute for Frontier Materials, Deakin University, Geelong Waurn Ponds Campus, Waurn Ponds, Victoria 3216, Australia
E-mail: luhua.li@deakin.edu.au; ian.chen@deakin.edu.au
Dr. R. Jones
Department of Physics, La Trobe University, Bundoora, Victoria 3086, Australia

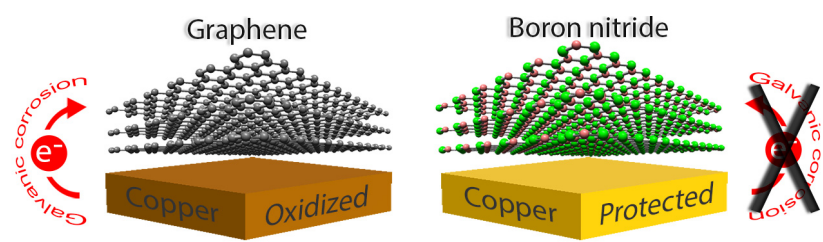

Although the high impermeability of graphene makes it an excellent barrier to inhibit metal oxidation and corrosion, graphene can form a galvanic cell with the underlying metal that promotes corrosion of the metal in the long term. Boron nitride (BN) nanosheets which have a similar impermeability could be a better choice as protective barrier, because they are more thermally and chemically stable than graphene and, more importantly, do not cause galvanic corrosion due to their electrical insulation. In this study, the performance of commercially available BN nanosheets grown by chemical vapor deposition as a protective coating on metal has been investigated. The heating of the copper foil covered with the BN nanosheet at 250 °C in air over 100 h results in dramatically less oxidation than the bare copper foil heated for 2 h under the same conditions. The electrochemical analyses reveal that the BN nanosheet coating can increase open circuit potential and possibly reduce oxidation of the underlying copper foil in sodium chloride solution. These results indicate that BN nanosheets are a good candidate for oxidation and corrosion protection, although conductive atomic force microscopy analyses show that the effectiveness of the protection relies on the quality of BN nanosheets.

## 1. Introduction

The deterioration of metal by oxidation and corrosion affects all aspects of industry, costing the United States around $300 billion each year.[1] Surface coatings using metals (including alloys), ceramics and organics to slow ion movement and hence corrosion rate is a common corrosion-control strategy.[2] Recently, graphene, a two-dimensional (2D) nanomaterial, has also been proposed as an oxidation and corrosion barrier on metals.[3-5] Graphene has many desirable properties, including a high impermeability to gases (including helium) and moisture,[6] high thermal conductivity[7] and low reactivity to most chemicals. Graphene coatings are also very thin, which minimizes geometric and morphological changes in the protected surface. Furthermore, it is almost transparent to visible light,[8] and hence causes little color change to the coated metal. A fatal problem with graphene, however, is that it can form a galvanic cell with the underlying metal and even accelerate oxidation and corrosion in the long term.[9,10]

Boron nitride (BN) nanosheets, sometimes called "white graphene", may provide a better alternative to graphene: one with the desirable properties of 2D nanomaterials, but without causing galvanic corrosion. BN nanosheets have comparable impermeability, mechanical properties and thermal conductivity to graphene.[11-17] In addition, BN nanosheets are more transparent to visible light due to their wide bandgap and have greater chemical and thermal stability.[18] Most importantly, BN nanosheets are electrically insulating[19] and do not enhance galvanic corrosion of the underlying metal. BN nanosheets of relatively large sizes can be produced by chemical vapor deposition (CVD).[11,20-27] As is the case for graphene, these can be used directly on the metal substrate or transferred to arbitrary substrates as an anti-corrosion coating. Recently, it was reported that a ~5 nm thick BN nanosheet could protect copper from oxidation at 500 °C for 0.5 h.[27] In fact, many practical applications require anti-oxidation protection of metals at relatively low temperatures over a long period of time. In this work, the performance of a commercially available CVD-grown BN nanosheet in inhibiting metal oxidation at 250 °C for up to 100 h and corrosion in sodium chloride (NaCl) solution has been examined by using electrochemical tests and analysing the effects of heating on coated and uncoated copper foils.

## 2. Results and Discussion

A bare Cu foil and one coated with CVD-grown BN nanosheets (BN+Cu) were used as received. The thickness of the BN nanosheet was determined to be 7-8 nm (~20 layers) by atomic force microscopy (AFM) after a transfer to a silicon substrate with a 90 nm oxide layer ($SiO_2$/Si) (see Supporting Information, Figure S1).[28] Thanks to BN's low absorption of visible light, the BN-covered Cu foil has a very similar metallic color to the bare Cu foil, as shown by the optical microscopy and digital camera photos in **Figure 1a** and 1e. The heating treatments were conducted at 250 °C in open air for up to 100 h, because oxidation resistance at a low to moderate temperature range for an extended period of time can mimic more the practical situations. The heating of the two foils produced contrasting color changes, indicating differences in their oxidation levels. The bare Cu foil turned dark-brown after just 2 h of heating (Figure 1b), as a result of the formation of large amounts of black cupric oxide (CuO) and red cuprous oxide ($Cu_2O$) on its surface.[3,9,10] Under the same conditions, the BN+Cu foil retained much of its metallic luster, indicating a much lesser degree of surface oxidation, although some black oxidized areas were observed under an optical microscope (Figure 1f). After heating from 20 to 100 h in air, the BN+Cu foil became light orange (Figure 1g and 1h), while the bare Cu foil became completely black (Figure 1c and 1d). The black regions observed on the BN+Cu foil under optical microscopy increased only slightly in size, from about 30% from 2 h heating to about 37% as the heating time increased to 100 h.

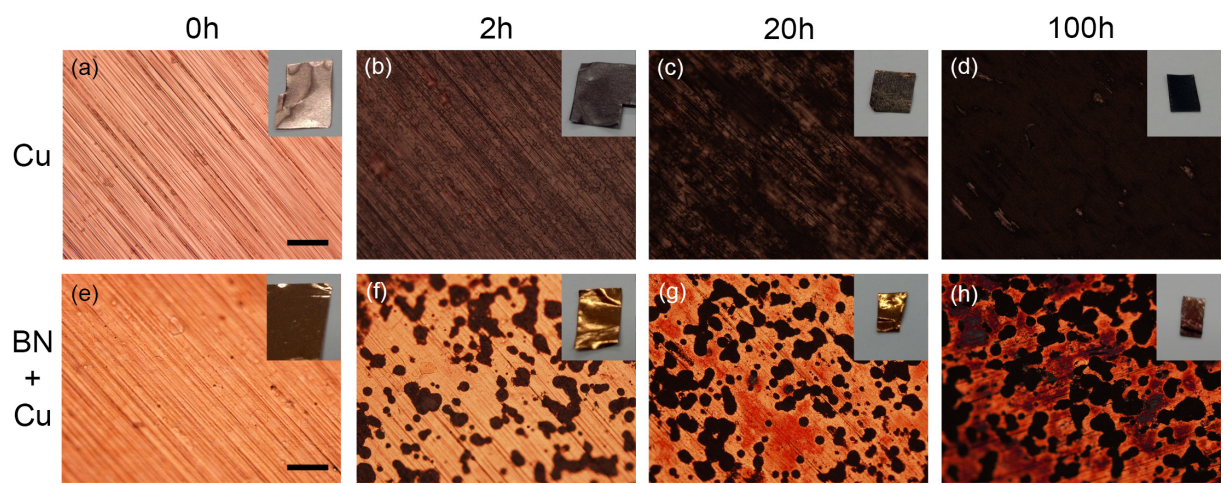

Figure 1. Optical microscopy photos of the bare Cu foil (Cu) and the BN-covered Cu foil (BN+Cu) before (0 h) and after heating at 250 °C in air for 2, 20 and 100 h. For comparison, the exposure time for all the photos is the same. The scale bars are 100 μm. The insets are the corresponding digital camera photos.

The surface morphology of the two foils after heating was examined by scanning electron microscopy (SEM). The bare Cu foil was completely covered with oxide particles after 2 h heating, with the size of these particles increasing as the heating time increased to 100 h (Figure 2b and c). In contrast, most of the surface of the BN+Cu foil was unchanged after heating for 2 h (Figure 2e) and decorated with only a small amount of oxide particles after 100 h (Figure 2f). Furthermore, the oxide particles on the BN+Cu foil after oxidation for 100 h (Figure 2f) are much smaller than those on the corresponding bare Cu foil (Figure 2b), indicating that the degree of oxidation is substantially reduced. This is confirmed by energy-dispersive X-ray (EDX) spectroscopy, which shows a strong intensity increase in the O signal of the bare Cu foil after 2 h heating, in contrast to the relatively low level for the BN+Cu foil even after the 100 h (Figure 3a and 3b). The atomic ratios between O and Cu from standardless quantitative analyses are plotted in Figure 3c. The O:Cu ratios for the unheated bare Cu and BN+Cu foils are similar (0.021±0.005 and 0.024±0.004, respectively), whereas the O:Cu ratio for the BN+Cu foil heated for 100 h (0.088±0.020) is substantially smaller than that of the bare foil heated for only 2 h (0.554±0.095).

As the EDX technique is not sufficiently surface sensitive to probe the BN nanosheet, chemical changes to the BN nanosheet before and after the 100 h heating treatment were investigated using X-ray photoelectron spectroscopy (XPS). The B 1s spectrum of the BN nanosheet on the copper foil prior to heating appears to comprise two components (Figure 4a): a dominant peak centered at 190.4 eV which is characteristic of B–N bonds[29] and a shoulder at 189.0 eV assigned to hydrogenated boron atoms (B–H)[30] which are intermediate products during the CVD growth process.[25] The N 1s spectrum of the BN nanosheet prior to heating is also dominated by the B-N bonds with the binding energy of 398.0 eV, along with a shoulder at 396.8 eV which is possibly associated with nitrogen atoms bonded to the hydrogenated boron atoms (Figure 4b).[30] After heating for 100 h, the shoulders at the lower binding energies in both the B 1s and N 1s spectra disappears. The N 1s spectrum reduces to a single peak corresponding to B–N; the B 1s spectrum, on the other hand, contains a new shoulder at 191.7 eV, which is attributed to the formation of B–O bonds.[29,31] The XPS results suggest that the BN nanosheet on the copper foil is not affected much by the heating. So it can be summarized that the BN nanosheet has an excellent protection of the underlying metal from oxidation at an elevated temperature for relatively long time.

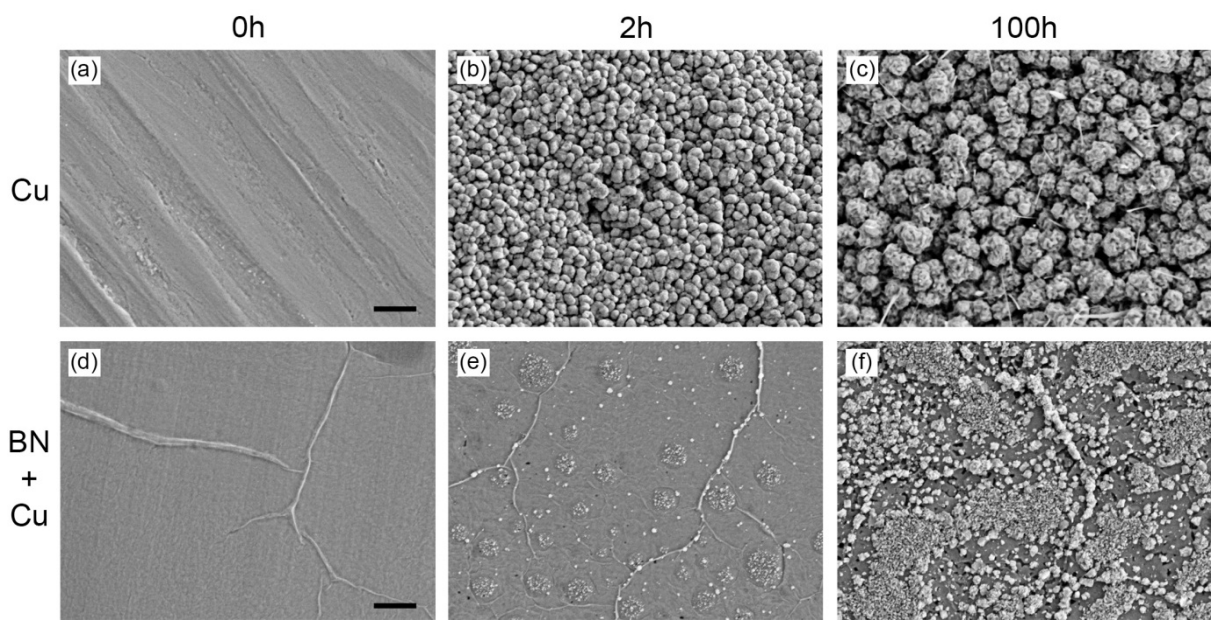

Figure 2. SEM images of the bare foil (Cu) and the BN covered Cu foil (BN+Cu) before and after heating at 250 °C in air for 2 and 100 h. All the images have the same magnification and the scale bars are 500 nm.

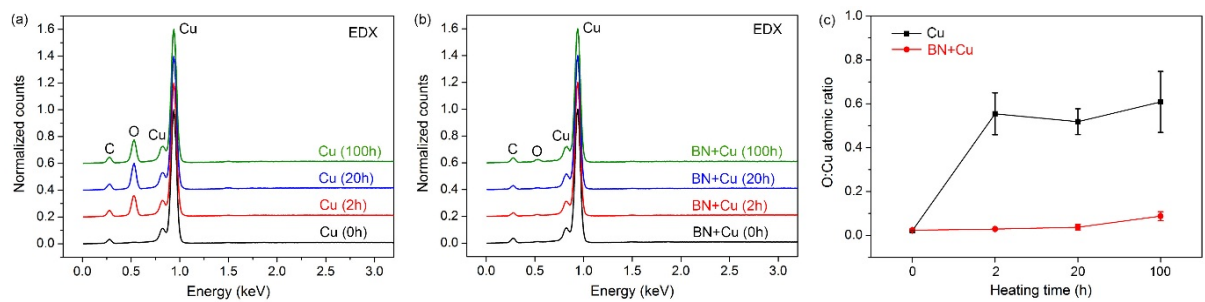

Figure 3. EDX spectra of (a) bare Cu foil (Cu) and (b) BN covered Cu foil (BN+Cu) after heating at 250 °C for different periods of time (0-100 h). (c) The standardless quantitative atomic ratios between O and Cu from the EDX spectra of the two foils.

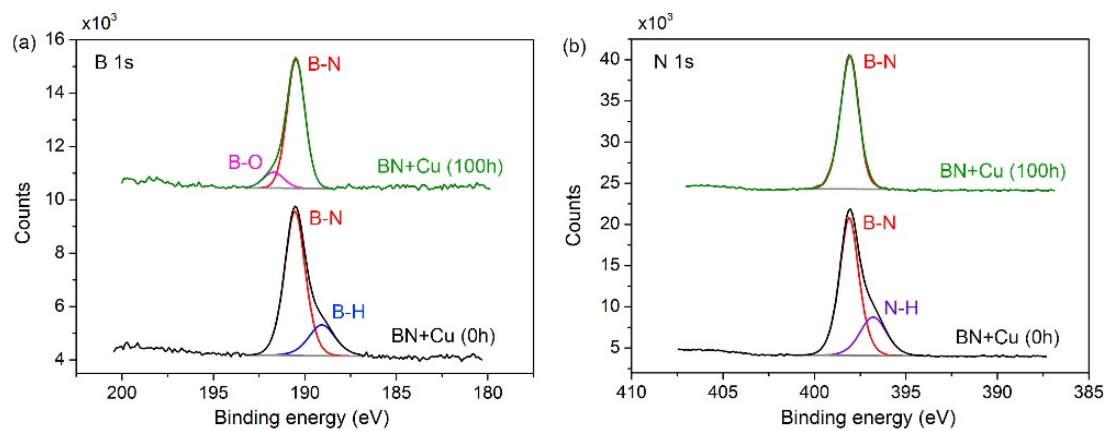

Figure 4. XPS spectra in (a) B 1s region and (b) N 1s region of the BN covered Cu foil (BN+Cu) before and after heating at 250 °C in air for 100 h.

Although the BN nanosheet shows remarkable performance in inhibiting oxidation of the Cu foil, its protection depends on the quality. As shown in the optical microscope images (Figure 1f-h), heating still resulted in the oxidation of about 30% of the surface of the BN+Cu foil. The SEM images in Figure 5 also show patches of oxide particles on the BN+Cu foil after heating for 2 h (white patches and white particles). The quality and homogeneity of the BN nanosheet was investigated using Raman spectroscopy. Most of the BN nanosheet produced a very weak Raman G band at around 1370 cm$^{-1}$, corresponding to the $E_{2g}$ vibration mode in hBN; however, for some areas this band was absent (see Supporting Information, Figure S4). Although BN has a much lower Raman yield than graphite, high-quality BN monolayers can still produce a sharp G band.[18,32] So the Raman results imply that the BN nanosheet is not of high quality. To gain a better understanding on the spatial variation in the quality of the BN nanosheet, we turned to conductive atomic force microscope (AFM). BN nanosheets are a wide bandgap semiconductor and electrical insulating; however impurities as well as low crystallinity can increase its conductivity.[33] Consequently, the variation in crystal quality can be determined from the local electrical conductivity by conductive AFM. Both the height image (Figure 5c) and the deflection image (Figure 5d) of the pristine BN+Cu foil (without heating) show some disk-like protrusions (arrows) and ripples (rectangle) on the surface of the BN nanosheet. Raman spectroscopy shows that the disk-like protrusions are BN of greater thickness (see Supporting Information, Figure S5). The conductivity map (Figure 5e) shows that the middle area of the BN nanosheet is insulating (dark red area representing currents of 0 nA) and that the upper-right and lower-left regions are much more electrically conducting (yellow area representing currents ≥8 nA). The overlay of the deflection and conductive AFM images reveals that the area around the disk-like protrusions is more conductive and hence of lower crystal quality (Figure 5f). This is very different from high-quality CVD-grown graphene, in which defects are mainly located at grain boundaries and edges.[3] It can be seen from the SEM image of the BN+Cu foil heated for 2 h (Figure 5b) that the oxide particles appear around the disk-like BN protrusions (arrows), whereas the area containing ripples of BN nanosheet is free of oxide particles. The excellent match between the conductive AFM image and the SEM image shows that the oxidation of the underlying Cu foil mainly happens at places where the BN nanosheet is of low quality, allowing oxygen to penetrate and react with the underlying Cu during heating. In addition, thicker BN tends to have better protection. The formation of B–O bonds shown in the XPS (Figure 4a) is likely to have occurred at these low-quality regions where the BN nanosheet could be torn up by the growth of the copper oxide particles and then oxidized, because BN nanosheets are found to be resistant to oxidation below 850 °C[18] and the heating of the transferred BN nanosheet on $SiO_2$/Si causes no morphology change (see Supporting Information, Figure S6). In contrast, the part covered by the BN nanosheet of a relatively good quality effectively blocks oxygen diffusion so that the underlying Cu is only slightly oxidized after heating for 100 h. These results suggest that the protection provided by BN nanosheets can depend on crystal quality.

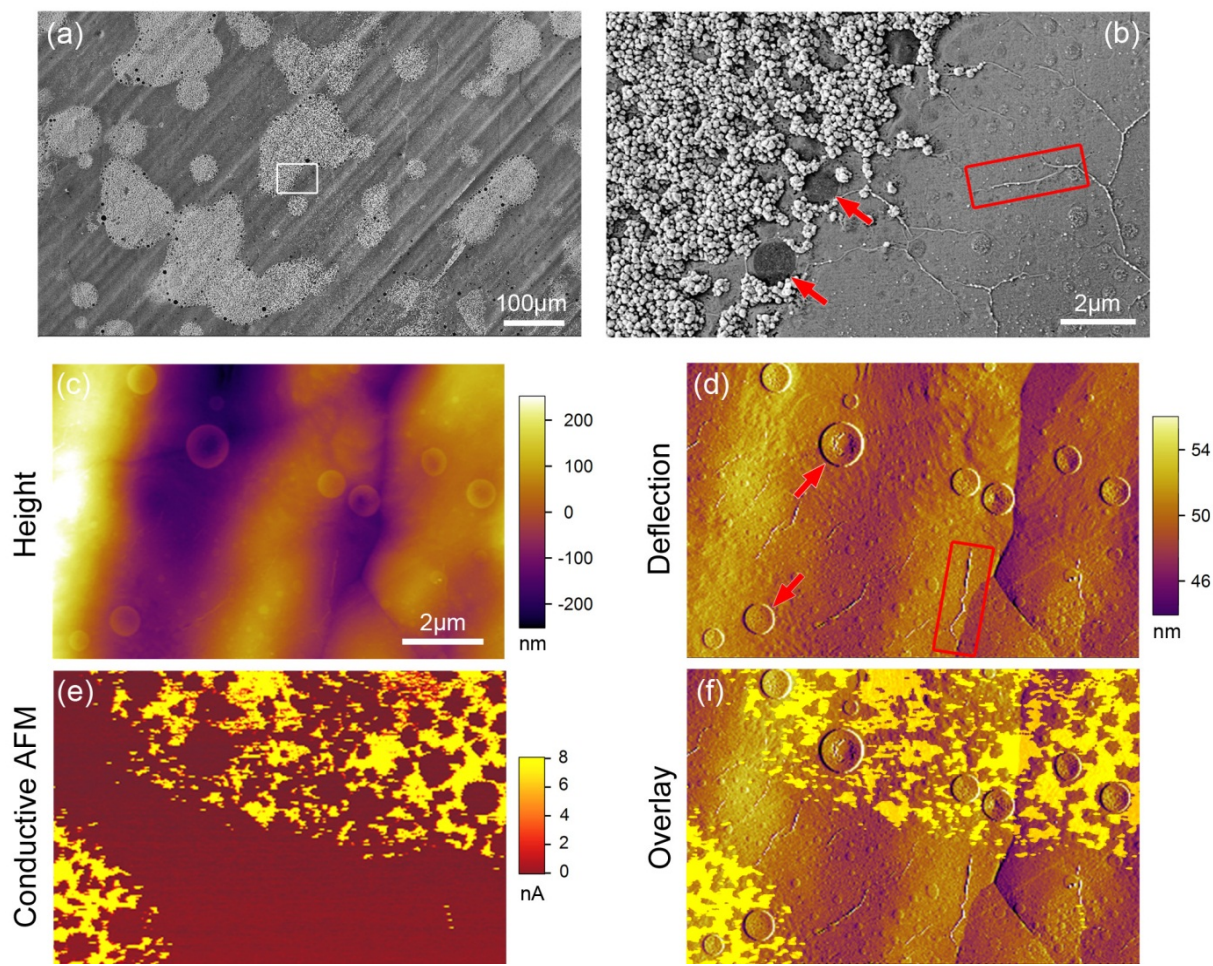

Figure 5. (a), (b) SEM images of the BN nanosheet covered Cu foil (BN+Cu) after the heating at 250 °C in air for 2 h; (b) magnified view of the rectangular area in (a); conductive AFM images of the BN+Cu foil without heating: (c) height image; (d) deflection image; (e) conductive AFM image under a bias voltage of 1 V and (f) overlay of the deflection and conductive AFM images. All the AFM images have the same scale bar.

The corrosion passivation of the BN nanosheet in aqueous environment was tested in 0.1 M NaCl solution by electrochemical methods. The open circuit potentials (OCPs) of the two foils show that the BN+Cu foil is nobler than the bare Cu foil. The BN+Cu foil reaches equilibrium very quickly, and the corrosion potential is about -152 mV (BN+Cu in Figure 6). In contrast, the OCP of the bare Cu foil initially drops off from -130 to -180 mV, probably due to the desolution of copper oxide on its surface,[34] and then gradually stabilizes at about -165 mV (Cu in Figure 6).

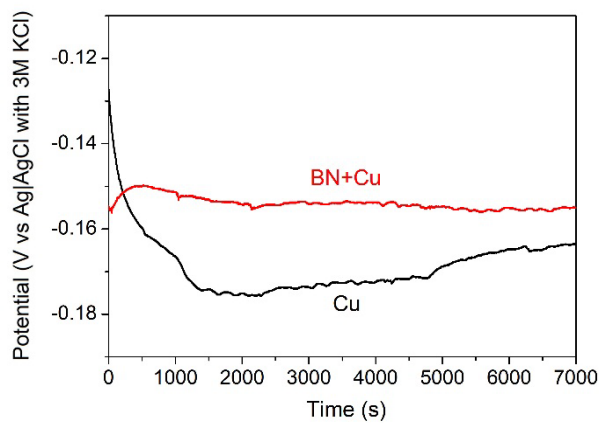

Figure 6. OCPs of the bare Cu foil (Cu) and BN covered Cu foil (BN+Cu) in 0.1 M NaCl solution.

Tafel analyses have also been carried out to illustrate the corrosion kinetics of the Cu foil with and without the BN nanosheet. In aerated NaCl solution, the BN+Cu foil shows a lower anodic current, but a higher cathodic current, than the bare Cu foil (Figure 7a), suggesting that the BN nanosheet deactivates copper oxidation but enhances oxygen reduction.[35] To confirm this, similar cyclic voltammetry (CV) tests were conducted in nitrogen-saturated NaCl solution. According to the Tafel plots, the BN+Cu foil has lower anodic and cathodic currents, and a decreased corrosion current density (Figure 7b). The corrosion currents ($I_{corr}$) of the bare Cu and the BN+Cu foils are 0.93 and 0.27 $\mu A/cm^2$, respectively. In addition, the corrosion potential shift for the BN+Cu foil from -0.20 V to about -0.17 V indicates a passivation of the underlying Cu by the nanosheet. The comparison between the Tafel plots from the aerated and nitrogen-saturated solution confirms that the higher cathodic current from the BN+Cu foil is due to the enhanced reduction of oxygen which may be due to pure BN nanosheets[36] or defective BN nanosheets. Carbon and oxygen impurities are often present in BN crystals[16,37] and carbon doping can form boron carbide nitride which has been found to be efficient in oxygen reduction.[38,39] Nevertheless, the lower anodic current from the BN+Cu foil in aerated solution and the excellent anti-oxidation performance in air suggests that the oxygen reduction by the BN nanosheet does not enhance the formation of $Cu_2O$ on the underlying Cu surface.

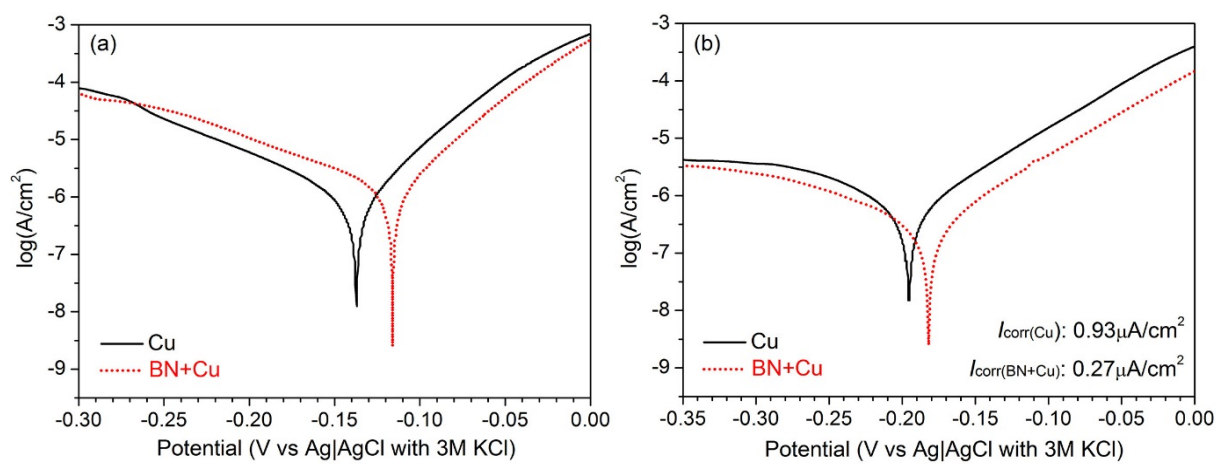

Figure 7. Tafel plots of the bare Cu foil (Cu) and BN nanosheet covered Cu foil (BN+Cu) in (a) aerated and (b) nitrogen gas bubbled 0.1 M NaCl solution.

## 3. Conclusions

The performance of a CVD-grown BN nanosheet in protecting Cu substrate from oxidation and corrosion has been investigated in details. The BN nanosheet effectively impedes oxidation of the underlying Cu. The Cu foil coated by the BN nanosheet heated at 250 °C in air for 100 h shows a dramatically lesser degree of oxidation than the bare Cu foil (without BN) heated for 2 h under the same conditions. The dynamics and kinetics of the corrosion passivation by the BN nanosheet in NaCl solution was revealed by OCP and Tafel analyses. The BN-covered Cu foil has a higher OCP and lower anodic current than the bare Cu foil. These results suggest that BN nanosheets are good candidates as barriers to oxidation and corrosion in metals and their effectiveness depends on their crystal quality.

## 4. Experimental Section

Both the Cu foil coated with CVD-grown BN nanosheet and the bare Cu foil (20 μm thick) were purchased from Graphene Supermarket and used as received. The BN nanosheets grown from borazine cover both sides of the copper foil. Oxidation of the foils was conducted in open air using a tube furnace (Tetlow). An Olympus BX51 optical microscope equipped with a DP71 camera and a Carl Zeiss Supra 55VP SEM equipped with an Oxford INCA EDX system were used to examine the foils before and after heating. X-ray photoelectron spectra were acquired using a Kratos AXIS Nova spectrometer (Kratos Analytical Ltd, U.K.) equipped with a monochromated Al Kα X-ray source ($hv$ = 1486.6 eV) operating at 150 W. The spectra were recorded at 0.1 eV/step and a pass energy of 20 eV. The pressure in the analysis chamber was below $4 \times 10^{-8}$ torr. The conductive AFM images were taken in contact mode under a bias voltage of 1 V using an Asylum Research Cypher scanning probe microscope (SPM) with a cantilever of conductive titanium and platinum (Ti/Pt) coating (Olympus AC240TM). All the electrochemical experiments were performed on an Ivium-n-Stat electrochemical analyser using a 3-electrode system, which had Ag|AgCl (3M KCl) as a reference electrode and Pt wire as a counter electrode. The electrolyte was prepared with NaCl (purity 99.9%, Univar) in deionised water at a concentration of 0.1 M. All the foils were cut to stripes with a width of 10 mm. During the electrochemical measurements, the foil strips of 4 mm were immersed in 40 mL electrolyte (with an immersed area of 40 mm$^2$). The nitrogen-saturated NaCl solution was prepared by bubbling with nitrogen gas (purity 99.99%, Coregas) for 20 mins before the measurements. The Tafel plots were acquired from CV tests with a scan rate of 1 mV/s from 0 to -0.4 V. The $I_{corr}$ were determined from the intersection of the linearly fitted anodic and cathodic currents.

**Supporting Information**
Supporting Information is available from the Wiley Online Library or from the author.

**Acknowledgements**
LH Li thanks the financial support from the CRGS2013 grant in Deakin University.

# Supporting Information

## Boron Nitride Nanosheets for Metal Protection

*Lu Hua Li,[1]\* Xing Tan,[1] Ying Chen[1]\* and Rob Jones[2]*

1. Institute for Frontier Materials, Deakin University, Geelong Waurn Ponds Campus, Waurn Ponds, Victoria 3216, Australia

2. Department of Physics, La Trobe University, Bundoora, Victoria 3086, Australia

\*Corresponding author: luhua.li@deakin.edu.au; ian.chen@deakin.edu.au

## 1. Thickness of the BN nanosheet

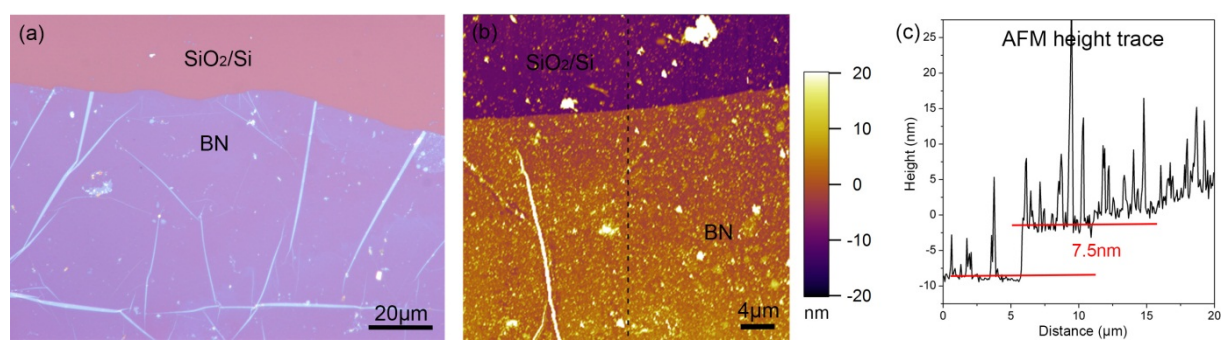

Figure S1. (a) Optical microscopy photo of the BN nanosheet after transferred to 90 nm SiO$_2$/Si substrate; (b) AFM image of the BN nanosheet on SiO$_2$/Si; (c) height trace of the dashed line in the AFM image.

The BN nanosheet was transferred to Si substrate covered with a layer of 90 nm thick oxide (SiO$_2$/Si) so that its thickness can be measured by atomic force microscope (AFM). In the transfer process, the BN nanosheet was first coated with a layer of polymethyl methacrylate (PMMA) (average Mw ~120,000, Sigma-Aldrich) which was dissolved in acetone. The Cu foil was then etched off using iron chloride ($FeCl_3$) solution. The BN nanosheet with PMMA was transferred to $SiO_2$/Si substrate in water. Finally, the PMMA was dissolved in acetone. Figure S1a shows an optical microscopy photo of the BN nanosheet on $SiO_2$/Si substrate. A Cypher AFM (Asylum Research) operated in tapping mode was utilized to measure the thickness of the BN nanosheet on $SiO_2$/Si using a Si cantilever with a spring constant of 7.4 N/m (NanoWorld). The AFM results show that the BN nanosheet is 7-8 nm thick (Figure S1b and c).

## 2. XPS analyses of the oxides

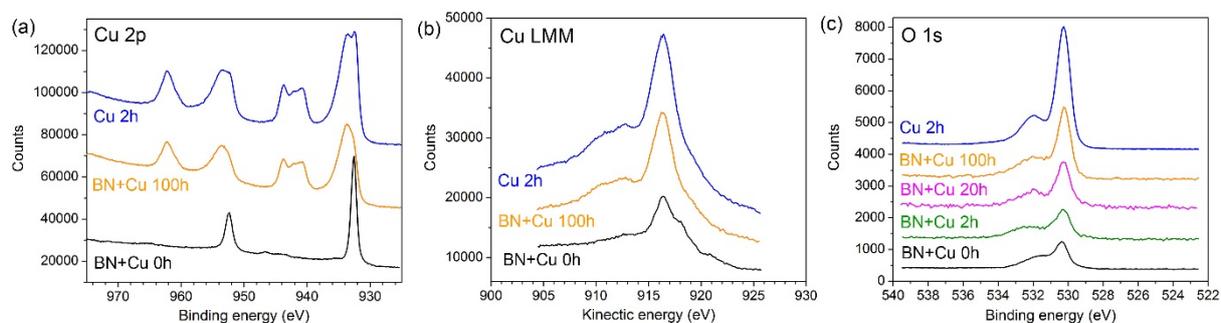

Figure S2. XPS spectra in (a) Cu 2p region; (b) Cu LMM region; (c) O 1s region, of the bare Cu foil (Cu) and BN covered Cu foil (BN+Cu) before and after different hours of heating at 250 °C in air.

Prior to heating, the principal component of the Cu 2p doublet (Cu $2p_{3/2}$) for the BN nanosheet covered Cu foil appears as a single peak at binding energy of 932.5 eV (Figure S2a), which is characteristic of both metallic copper ($Cu^0$) and copper(I) oxide ($Cu_2O$). To distinguish between these two states, it was necessary to also record the Cu LMM Auger emission spectrum, which is plotted on a kinetic-energy scale in Figure S2b. For the unheated sample, this comprises an intense peak at 916.4 eV, corresponding to $Cu^I$, and a shoulder near 918.0 eV, corresponding to $Cu^0$. From the relative intensities of these two components, it is concluded that most of the copper present on the surface is $Cu_2O$. After the heating at 250 °C for 100 h, the Cu $2p_{3/2}$ peak shifts to 933.5 eV and satellite peaks appear on the higher binding-energy side of this and the Cu $2p_{1/2}$ component. Both are indicative of oxidation to copper(II) oxide, CuO.

### 3. Black spots on the oxidized BN+Cu foil

The oxidation degrees of the black spots on the BN+Cu foil after heating are investigated using EDX. The O:Cu atomic ratios of the black spots after heating at 250 °C for 0-100 h are dramatically larger than those of the rest of the foil, indicating dramatically more oxidation.

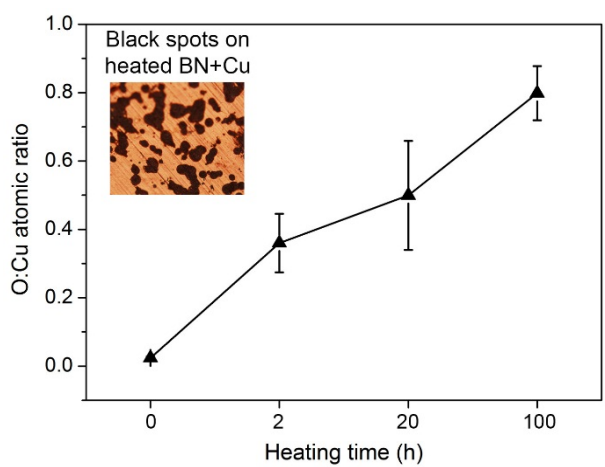

Figure S3. The standardless quantitative O:Cu atomic ratios from the EDX analyses on the black spots on the BN+Cu foil after different hours of heating.

### 4. Characterization of the multilayer BN nanosheets using Raman

A confocal micro-Raman spectrometer (inVia Renishaw) equipped with a 514.5 nm laser with a maximum power of 50 mW was used to characterize the BN nanosheet on Cu foil before any heating treatment. A 100x optical lens was used during the measurements. The BN nanosheet on Cu foil shows a very weak Raman G band at ~1370 cm$^{-1}$ (Figure S4b), indicating its crystal quality is not good. Though the mapping using the Raman intensity difference at 1370 (G band of BN) and 1420 (background) cm$^{-1}$ shows that the BN nanosheet has certain inhomogeneity in the crystal quality (Figure S4c), the weak G band and strong background signals may make the mapping not very accurate.

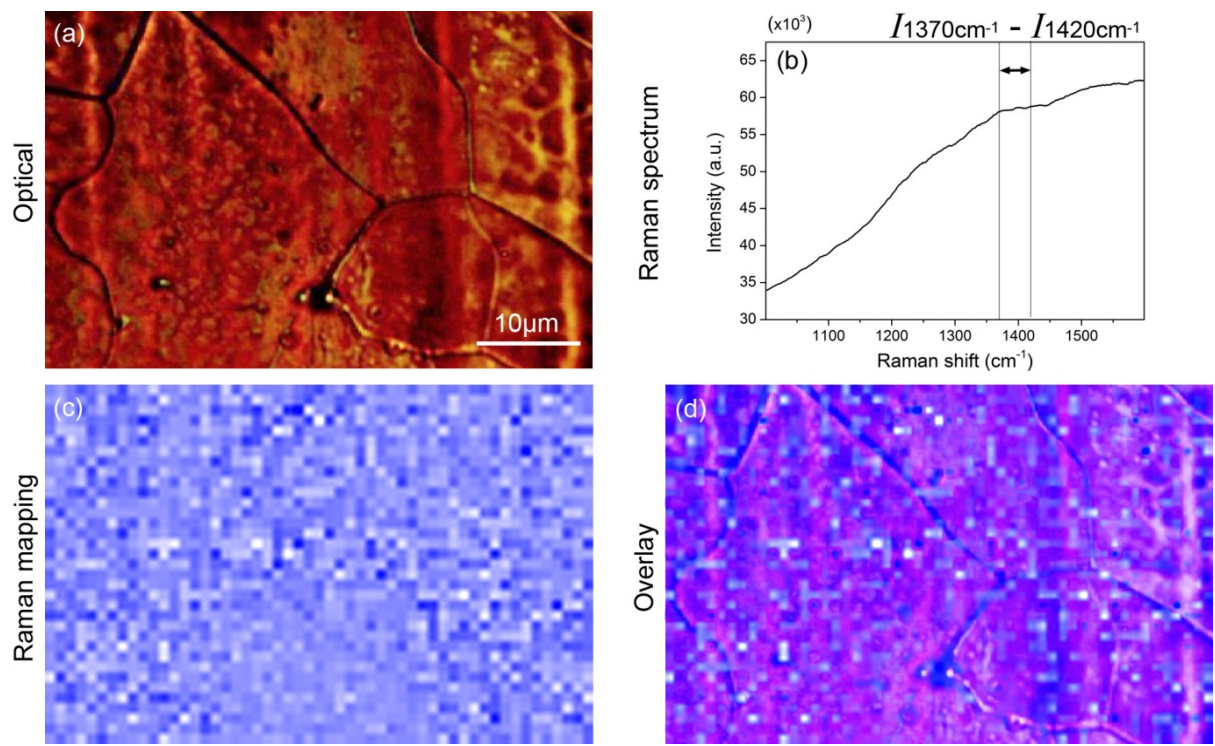

Figure S4. (a) Optical microscopy photo of the Cu foil covered with a layer of multilayer BN nanosheet; (b) a typical Raman spectrum of the BN nanosheet with a weak G band at ~1370 cm$^{-1}$; (c) Raman mapping of the area using the intensity difference at 1370 (G band of BN) and 1420 (background) cm$^{-1}$ and the brighter dots represent higher G band signal than the darker dots; (d) the overlay of the optical photo and Raman mapping.

## 5. Raman spectrum of the disk-like protrusions

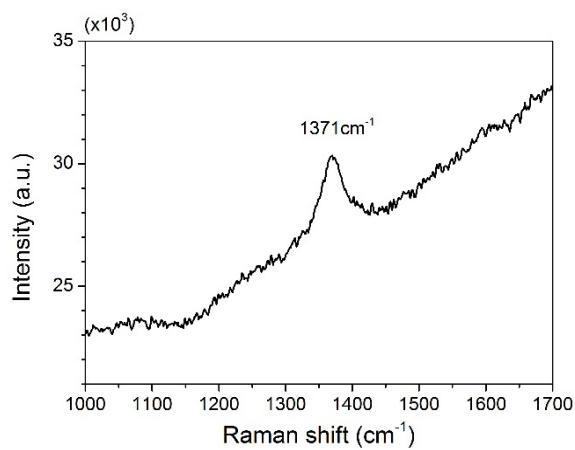

Figure S5. Raman spectrum from the disk-like protrusions.

The micro-Raman investigations show that the disk-like protrusions on the surface of the BN+Cu foil are thicker hBN (Figure S5).

## 6. Heating effect on the BN nanosheet

To test whether the BN nanosheet is affected by the heating, the BN nanosheet transferred to the 90 nm SiO2/Si substrate was heated at 250 °C in air for 2 h. According to both optical microscopy and AFM investigations, there is no noticeable morphology change of the nanosheet after the heating (Figure S6).

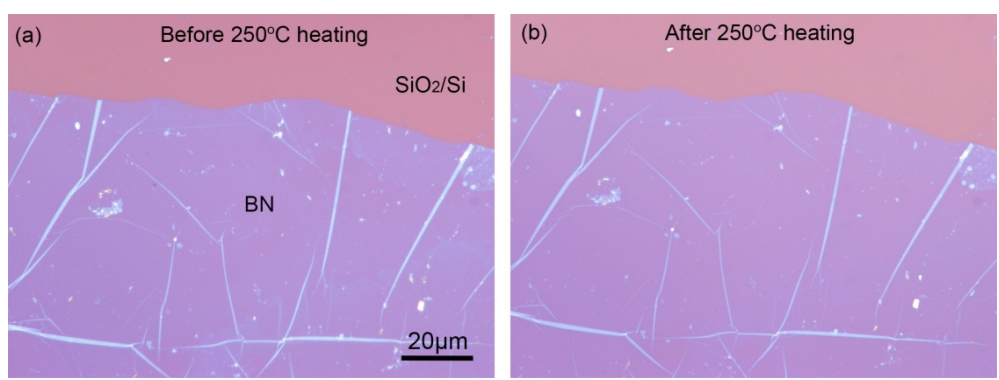

Figure S6. Optical microscope photos of the BN nanosheet on $SiO_2$/Si substrate (a) before any heat treatment; (b) after heating at 250 °C in air for 2 h.

## 7. Oxidation in air at 300 °C

The oxidation of the BN+Cu foil was also tested at 300 °C for up to 100 h. Figure S7 shows the optical photos of BN+Cu after heating in air at 300 °C for 20 and 100 h. It can be seen that the Cu was well protected after 20 h oxidation, but seriously oxidized after 100 h heating.

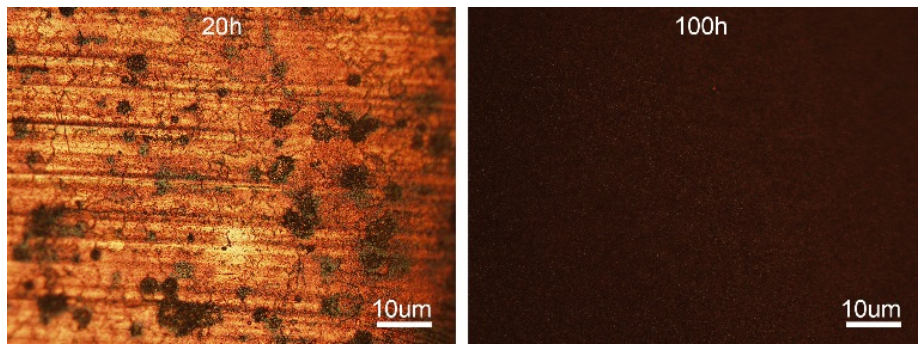

Figure S7. Optical photos of BN+Cu foil after heating in air at 300 °C for 20 and 100 h. For comparison, the same exposure time was used for both photos.